\documentclass[11pt,a4paper]{article}
\pdfoutput=1
\usepackage{jheppub}
\usepackage[utf8]{inputenc}
\usepackage{amsmath, amsthm, mathtools}
\usepackage{hyperref}
\usepackage{slashed}
\usepackage{color}
\usepackage{float}
\usepackage{xcolor}
\usepackage{graphicx}
\usepackage{xspace}
\usepackage{wrapfig,enumerate,slashed}
\usepackage[bottom]{footmisc}
\usepackage{wasysym} 
\usepackage{graphicx}
\usepackage{color}
\usepackage{orcidlink}
\usepackage{hyperref}
\usepackage{orcidlink}
\usepackage{subfigure}

\title{Harnessing Higgs Kinematics for HEFT Constraints}
\author[a]{Christoph Englert\orcidlink{0000-0003-2201-0667},}
\author[b]{Tom Ingebretsen Carlson\orcidlink{0000-0002-3699-8517},}
\author[b]{J\"orgen Sj\"olin\orcidlink{0000-0002-5285-8995},}
\author[c]{Michael Spannowsky\orcidlink{0000-0002-8362-0576}}
\affiliation[a]{School of Physics \& Astronomy, University of Glasgow, Glasgow G12 8QQ, UK}
\affiliation[b]{Department of Physics, Stockholm University, Sweden}
\affiliation[c]{Institute for Particle Physics Phenomenology, Department of Physics, Durham University, Durham DH1 3LE, UK}
\emailAdd{christoph.englert@glasgow.ac.uk}
\emailAdd{tom.ingebretsen-carlson@fysik.su.se}
\emailAdd{sjolin@fysik.su.se}
\emailAdd{michael.spannowsky@durham.ac.uk}
\abstract{We present a momentum-dependent reweighting strategy to extend current LHC di-Higgs analyses within the $\kappa$-framework and SMEFT into the bosonic sector of the Higgs Effective Field Theory (HEFT). Unlike SMEFT, where symmetry constraints tightly correlate multi-Higgs processes, HEFT allows for a broader range of momentum-dependent deviations that can substantially impact di-Higgs kinematics and offer a powerful probe of non-linear Higgs dynamics. We generalise the interpretation of existing experimental analyses by integrating HEFT operators up to chiral dimension four into differential Monte Carlo reweighting. We quantify the sensitivity to representative HEFT operators using multi-dimensional likelihoods for Run 3 and project the reach at the High-Luminosity LHC (HL-LHC). Particular emphasis is placed on how different exclusive final states, such as $b\bar{b}b\bar{b}$ and $b\bar{b}\gamma\gamma$, respond to momentum enhancements and how their complementary event selections drive exclusion limits. We further explore how rare final states, especially four-top production, can provide orthogonal constraints on HEFT-induced modifications, thereby enhancing global sensitivity to new physics effects in the Higgs sector.}

\keywords{}
\preprint{IPPP/25/32}
\begin{document}
\maketitle
\allowdisplaybreaks
\flushbottom
\section{Introduction}
As constraints on new physics from Higgs signal strength measurements and exotics' production are becoming increasingly tight at the Large Hadron Collider (LHC), limits on the production of multiple Higgs bosons are still relatively far away from sensitivity to the prediction of the Standard Model (SM). With largely SM-consistent measurements at the energy frontier of the LHC so far, this highlights di-Higgs production as a critical tool for the discovery of phenomena beyond the SM (BSM) in a phenomenological arena where deviations could still be sizable. The dominant gluon fusion mode $gg\to hh$ is also the most promising avenue to get a direct handle on the Higgs boson's self-coupling, which is understood to be one of the key parameters that could connect the electroweak scale to a more fundamental theory of particle interactions.

Modifications of di-Higgs final states are predicted in many concrete BSM scenarios where coupling modifications are typically accompanied by new resonance structures in the $hh$ final states (see e.g.~\cite{ATLAS:2019qdc,CMS:2015jal,CMS:2018tla,CMS:2022hgz}). Among a range of exclusive final states, the statistically most promising channels are $b\bar b b\bar b$ \cite{Dolan:2012rv, FerreiradeLima:2014qkf}, $ b\bar b \tau^+ \tau^-$ \cite{Barr:2013tda} and $b\bar b \gamma \gamma$~\cite{PhysRevD.100.096001}. Searches for resonant extensions of the SM \cite{Dolan:2012ac} are complemented by non-resonant effective field theory (EFT) analyses, especially applying SMEFT~\cite{Grzadkowski:2010es,Heinrich:2022huh}, which constructs BSM interactions from the SM particle content, in particular relying on operators derived from the Higgs doublet $\Phi$. In this `linear' framework, the value of Higgs pair production is somewhat limited; SM gauge symmetry tightly correlates the phenomenological implications across different Higgs boson multiplicities. For SM-consistent single-Higgs measurements, $gg\to hh$ dominantly singles out interactions $\sim (\Phi^\dagger \Phi)^3$ ($\Phi$ denotes the SM Higgs doublet), which manifests itself in the departure of the Higgs boson's trilinear coupling from its SM predicted value. (The latter is fixed by the electroweak vacuum expectation value (the $W$ mass) and the observed Higgs mass.) Put differently, `kappa' framework modifications~\cite{LHCHiggsCrossSectionWorkingGroup:2016ypw} of the Higgs trilinear interaction can be mapped transparently to SMEFT di-Higgs studies when single Higgs boson observables are consistent with the SM.

The expectation of these correlations has continuously shaped the experimental analyses program, principally through identifying particular signal regions of the mentioned exclusive $hh$ final states and their direct experimental value when viewed against rescaling of the Higgs self-coupling. As motivated as this approach might appear, given the current experimental outcome at the LHC, the imposed theoretical bias could distract from unexpected modifications of the Higgs sector, and this is in the very channel where deviations could be most telling. This may explain the growing interest in `non-linear' (or Higgs EFT, HEFT) effects in the Higgs sector (e.g.~\cite{Longhitano:1980tm,Feruglio:1992wf,Appelquist:1993ka,Alonso:2012px,Brivio:2013pma}) in the recent literature, especially because they provide unique opportunities for the LHC BSM Higgs physics agenda. HEFT highlights the Higgs boson's properties as a custodial singlet, a priori unrelated to the mechanism of electroweak symmetry breaking in a theoretically consistent manner. Several calculations~\cite{Delgado:2013hxa,Asiain:2021lch,Gavela:2014uta,Buchalla:2020kdh,Herrero:2021iqt,Anisha:2024ryj} have highlighted the flexibility and theoretical appeal of this framework in interpreting Higgs physics data and the correlation of Higgs measurements across different processes.
In this context, Higgs pair production is uniquely placed as the first phenomenological instance that could tension the Higgs multiplicity correlations predicted in SMEFT. Are these correlations experimentally observed? And if there is a deviation in the $hh$ pair production approaching the high luminosity (HL) LHC phase, can it clarify its linear vs. non-linear nature? This work aims to provide a step towards finding answers to these questions. We clarify how HEFT-like interactions highlight different exclusive channels to set constraints on momentum-dependent modifications of Higgs self-interactions that do not follow the dominant SMEFT paradigm. Capitalising on differential reweighting techniques~\cite{Cadamuro:2025car}, we demonstrate how bosonic HEFT interactions can be straightforwardly included in the existing experimental workflow. Without significant overhead, this enables the generalisation of the existing di-Higgs measurement effort towards a more comprehensive study of Higgs interactions that goes beyond the current coupling rescalings. In passing, we also highlight HEFT-specific correlations with other rare LHC final states that could clarify HEFT-like properties beyond Higgs pair production, with a focus on four top production, motivated by Higgs propagator considerations~\cite{Englert:2019zmt}.

This work is organised as follows. In section~\ref{sec:heft}, we provide details on di-Higgs production in the bosonic HEFT, up to chiral dimension four. We sketch an implementation of reweighting techniques~\cite{Cadamuro:2025car} that enables the experimental communities to carry out similar analyses with little overhead, using their existing Monte Carlo toolchains. In section~\ref{sec:lhc}, we turn to expected results by carefully reproducing existing LHC analyses and recasting them in the bosonic HEFT. Special attention is given to representative momentum-dependent enhancements not expected in dimension-six SMEFT. We also provide an extrapolation to the HL-LHC phase and highlight the relevance of additionally sensitive rare final states for HEFT-like analyses, specifically focusing on four top productions. We provide conclusions in section~\ref{sec:conc}.
\section{HEFT Higgs interactions for hh production}
\label{sec:heft}
HEFT interactions have been considered in a range of theoretical studies, in particular after the Higgs boson discovery~\cite{Brivio:2013pma,Buchalla:2013rka,Brivio:2016fzo}. Radiative corrections within this framework have been scrutinized~\cite{Delgado:2013hxa,Herrero:2021iqt,Sun:2022ssa,Anisha:2024ryj,Davila:2023fkk,Anisha:2022ctm,Herrero:2022krh,Anisha:2024ljc}. We will focus on Higgs pair production, for which phenomenological studies so far typically restrict interactions to the dominant chiral-dimension two operators. The relation with the SMEFT and, chiefly, the departure from these interactions in the di-Higgs mode are well-documented. In this work, we will present an extended analysis of di-Higgs production with a particular focus on how non-trivial HEFT momentum dependencies can sculpt the experimentally used paradigms that have been established through chiral dimension-2 studies. We aim to clarify the sensitivity potential that can be achieved when such interactions are present, and how experimental analyses beyond SMEFT correlations add value to LHC data interpretation across different channels, in particular in concert with other di-Higgs observation modes such as $pp \to t\bar t t \bar t$.

Within the bosonic HEFT, the three-point function that is typically associated with $\kappa_H$ rescalings receives corrections from a multitude of operators, in particular when chiral dimension four interactions are included (see e.g.~\cite{Herrero:2021iqt,Anisha:2024ljc} for overviews). In approaches tailored to experimental analyses based on momentum-dependent interpolations~\cite{Cadamuro:2025car}, changes to the Higgs two-point and three-point vertices, relevant for the {\emph{reducible}} 3-point function entering the $gg\to hh$ subamplitudes, can be obtained by a momentum-dependent replacement of the self-coupling
\begin{equation}
\parbox{3.8cm}{\includegraphics[height=2.3cm]{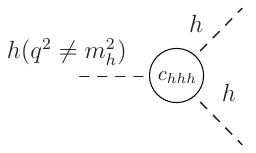}}
\to
\parbox{3.0cm}{\includegraphics[height=2.3cm]{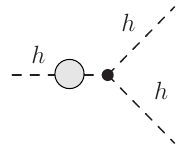}}
+
\parbox{3.0cm}{\includegraphics[height=2.3cm]{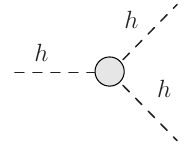}}\,,
\end{equation}
where the light-shaded regions denote the 1-PI irreducible insertions and we consider interactions up to chiral dimension four (in the on-shell scheme) without including radiative chiral dimension-two corrections (the impact of these have been comprehensively discussed in~\cite{Herrero:2022krh,Anisha:2024ljc}).\footnote{In the spirit of EFT each HEFT coefficient needs to be treated as an independent input parameter. Our results can therefore also be understood as a qualitative representation of the value of di-Higgs measurements and their non-standard momentum dependence in such a comprehensive program going beyond SMEFT.} Both vertex and propagator contributions, due to their scalar nature, can be effectively parametrised as a momentum-dependent redefinition
\begin{equation}
\label{eq:rescale}
\begin{split}
c^{\text{HEFT}}_{hhh,\chi 4} &= c_{hhh} \left( 1- \frac{e^2}{2 s_W^2} \frac{q^2+m_h^2}{m_W^2} a_{\Box\Box}\right) \\
&\hspace{1cm}+ \frac{e^2}{12 c_W^2s_W^2}
\frac{1}{m_h^2 m_W^2} \big\{
(q^2+2m_h^2)m_W^2 a_{ddZ} 
\\ 
&\hspace{2cm} + c_W^2 (q^4 a_{dd\Box} +2 m_h^2 [m_W^2 a_{ddW} + m_h^2 ( a_{h\Box\Box} + a_{hdd} )]
\\&
\hspace{2cm} + q^2 (m_W^2 a_{ddW} + m_h^2 [4 a_{h\Box\Box}+a_{hdd} -4a_{dd\Box }])) 
\big\},
\end{split}
\end{equation}
where on-shell kinematics for two Higgs legs have been enforced as required for the $gg\to hh$ process. Here, the various $a_i$ are the HEFT coefficients that arise due to non-minimal HEFT operators (the operator structures are explicitly listed in Tab.~1 of Ref.~\cite{Anisha:2024ljc}).

The only relevant scale is therefore the invariant di-Higgs mass $q^2=m^2_{hh}$. Note that within the bosonic HEFT, this modification, including the specific choice of kinematics, can be imported globally into higher-order QCD corrections~(see e.g. \cite{Heinrich:2022idm}) or LO jet-merged approaches without loss of generality.

Although the interactions arise from different HEFT coefficients $a_i$, as it becomes apparent from Eq.~\eqref{eq:rescale}, there are flat directions when their impact on the Higgs-trilinear self-coupling is considered. These cannot be disentangled via Higgs pair production alone. As we are predominantly interested in the impact of non-SM momentum dependences, we will investigate representative choices of $a_i$ that parametrise the different momentum dependencies. Most notably, $a_{\Box\Box}$ induces a new momentum dependence $\sim q^4$ into the Higgs self-energy, which can significantly shift the results away from the SM(EFT) expectation, see also \cite{Brivio:2014pfa,Anisha:2024xxc}. Further, $a_{dd\Box}$ is a prime candidate to introduce tell-tale momentum enhancements in the $hh$ final states. It is the combination of these operators that will enable us to track the importance of different experimental selections in setting HEFT limits in the future. \footnote{These operators are also the most important ones contributing to the structures of Eq.~\eqref{eq:rescale}.}

\section{Implications for the LHC}
\label{sec:lhc}
\subsection{Towards experimental analyses of HEFT $hh$ shapes}
\label{sec:details}
\begin{figure}[!t]
\begin{center}
\subfigure[]{\includegraphics[width=0.495\textwidth]{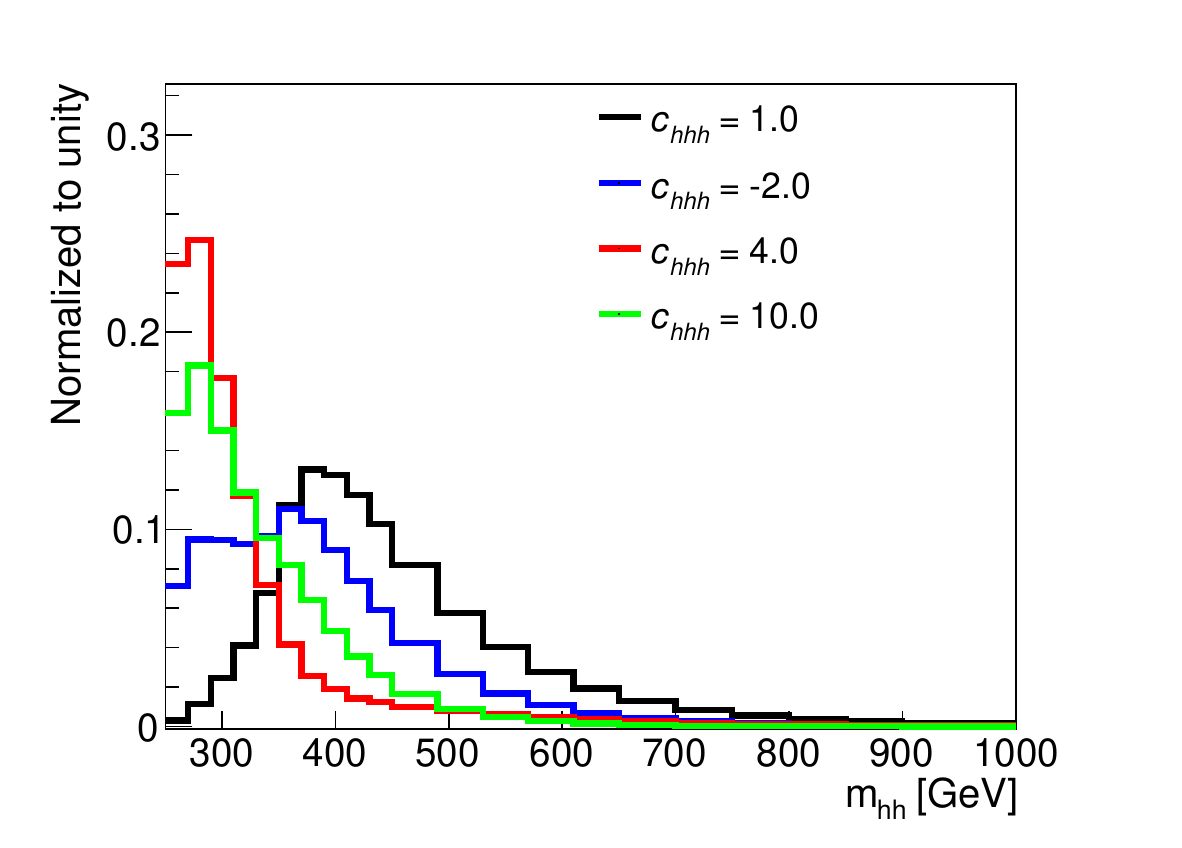}  }\hfill
\subfigure[]{\includegraphics[width=0.495\textwidth]{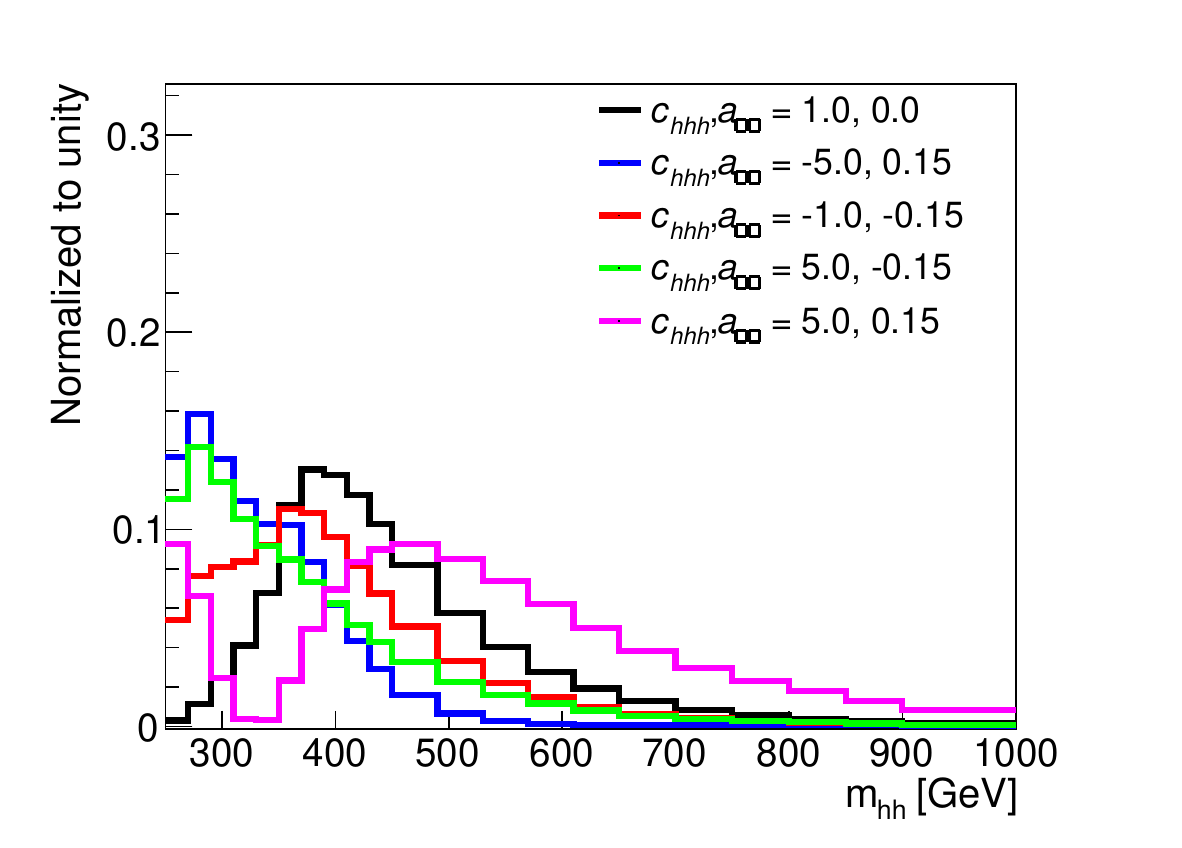}}
\subfigure[]{\includegraphics[width=0.495\textwidth]{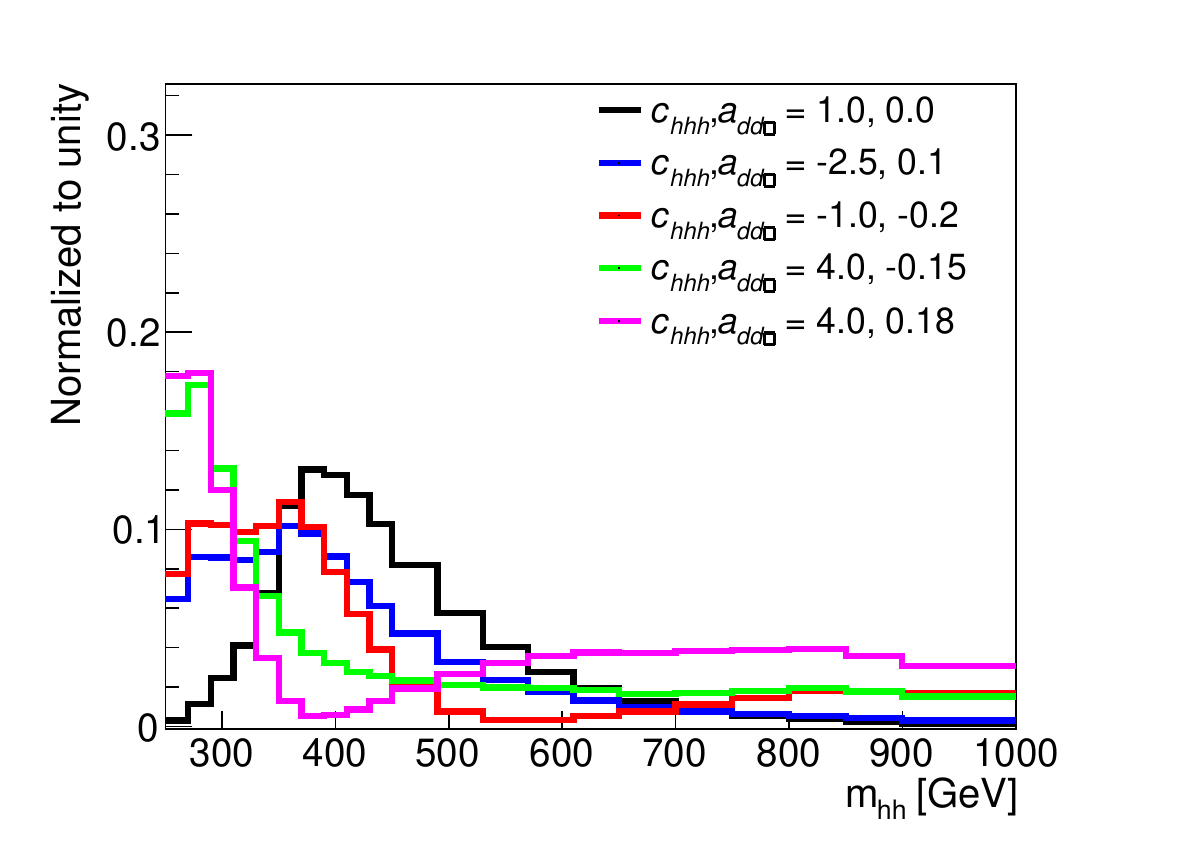} }
\hfill
\parbox{0.488\textwidth}{
\vspace{-2cm}
\caption{Representative distributions for different values of the (phenomenology-dominating) HEFT coefficients of Eq.~\eqref{eq:rescale}, demonstrating that a wide range of momentum-dependent shape modifications of the crucial $m_{hh}$ distribution can be achieved.\label{fig:mhhdists}}}
\end{center}
\end{figure}

The momentum dependencies introduced by HEFT operators are capable of sculpting the $m_{hh}$ distribution in ways that extend beyond, e.g., the benchmarking that has been performed in the context of the Higgs Cross Section Working Group~\cite{LHCHiggsCrossSectionWorkingGroup:2016ypw}. Representative shape examples are provided in Fig.~\ref{fig:mhhdists}, demonstrating a wide range of phenomenological outcomes that should predominantly inform the following discussion rather than be considered realistic HEFT parameter choices from the point of view of perturbativity.

As a proxy for more inclusive selection, we focus in the following on the $b\bar b \gamma \gamma$ mode, which we contrast with $b\bar b b \bar b$. $b\bar b b\bar b$ relies on harder selections to combat backgrounds and is therefore a good candidate to highlight correlation changes that become relevant in HEFT. Throughout, we focus on gluon fusion and neglect weak boson fusion (for a recent analysis in the context of HEFT, we refer the reader to~\cite{Herrero:2022krh,Anisha:2024ryj,Jager:2025isz,Braun:2025hvr}).

As a $2\to 2$ process, the hard scattering of $gg\to hh$ is described by the Mandelstam parameters $s$ (the invariant di-Higgs mass $m_{hh}$) and $t$ (scattering angle). The HEFT interactions considered here dominantly modify the $s$-dependence; however, the interference with box Feynman topologies can imply non-trivial, correlated $t$-sensitivity. Close to the threshold, angular information that relates to $t$ does not give rise to additional discriminating sensitivity, Fig.~\ref{fig:cosdistsHM}, but angular dependencies become more relevant for larger invariant masses, Fig.~\ref{fig:cosdistsHM}, and the $b\bar b b\bar b$ channel that access this boosted regime by construction.

\begin{figure}[!t]
\begin{center}
\subfigure[]{\includegraphics[width=0.496\textwidth]{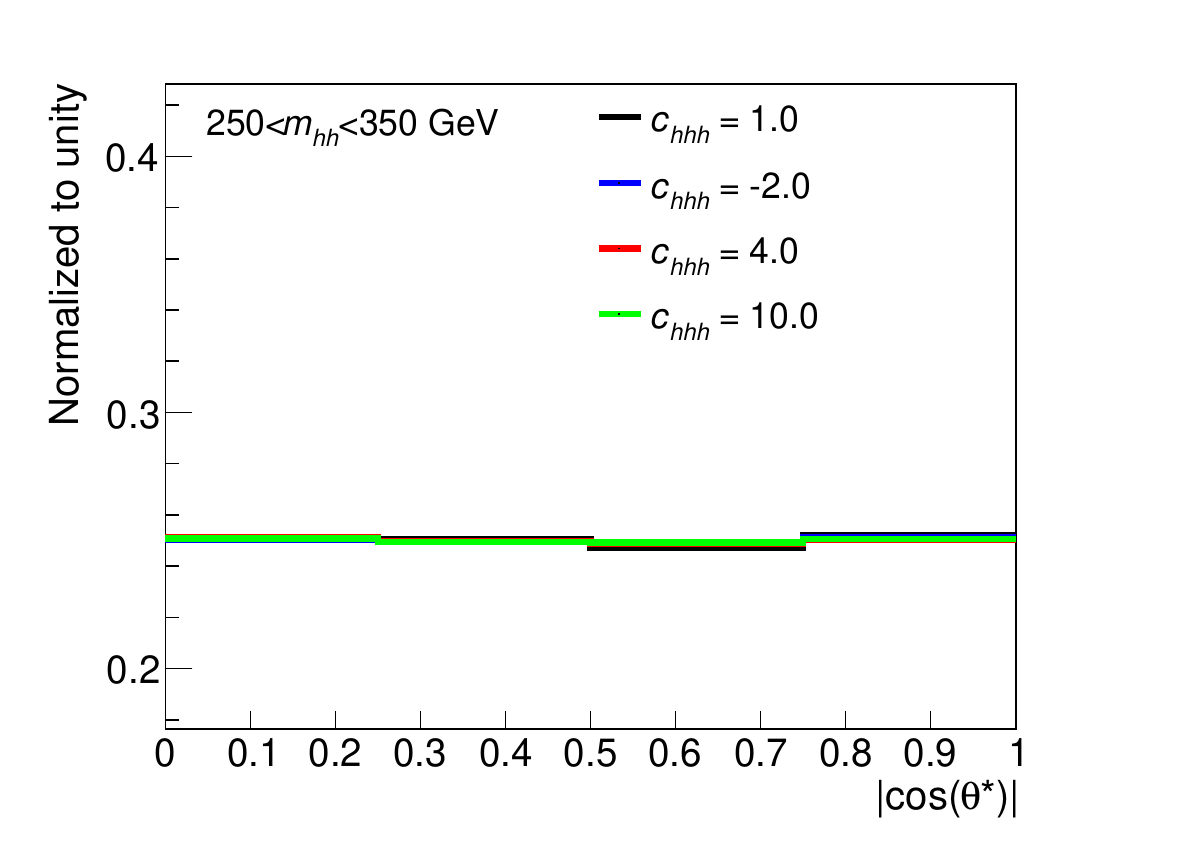}}\hfill
\subfigure[]{\includegraphics[width=0.496\textwidth]{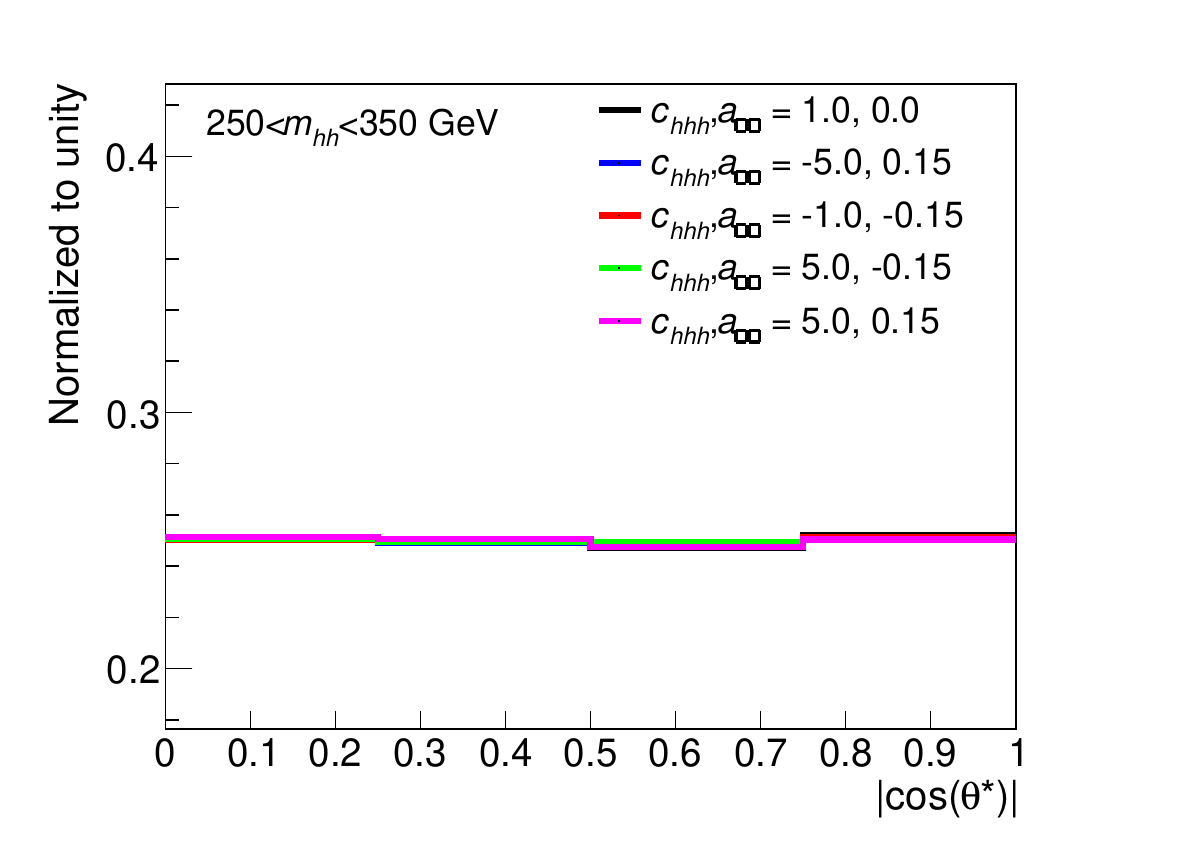}}
\subfigure[]{\includegraphics[width=0.496\textwidth]{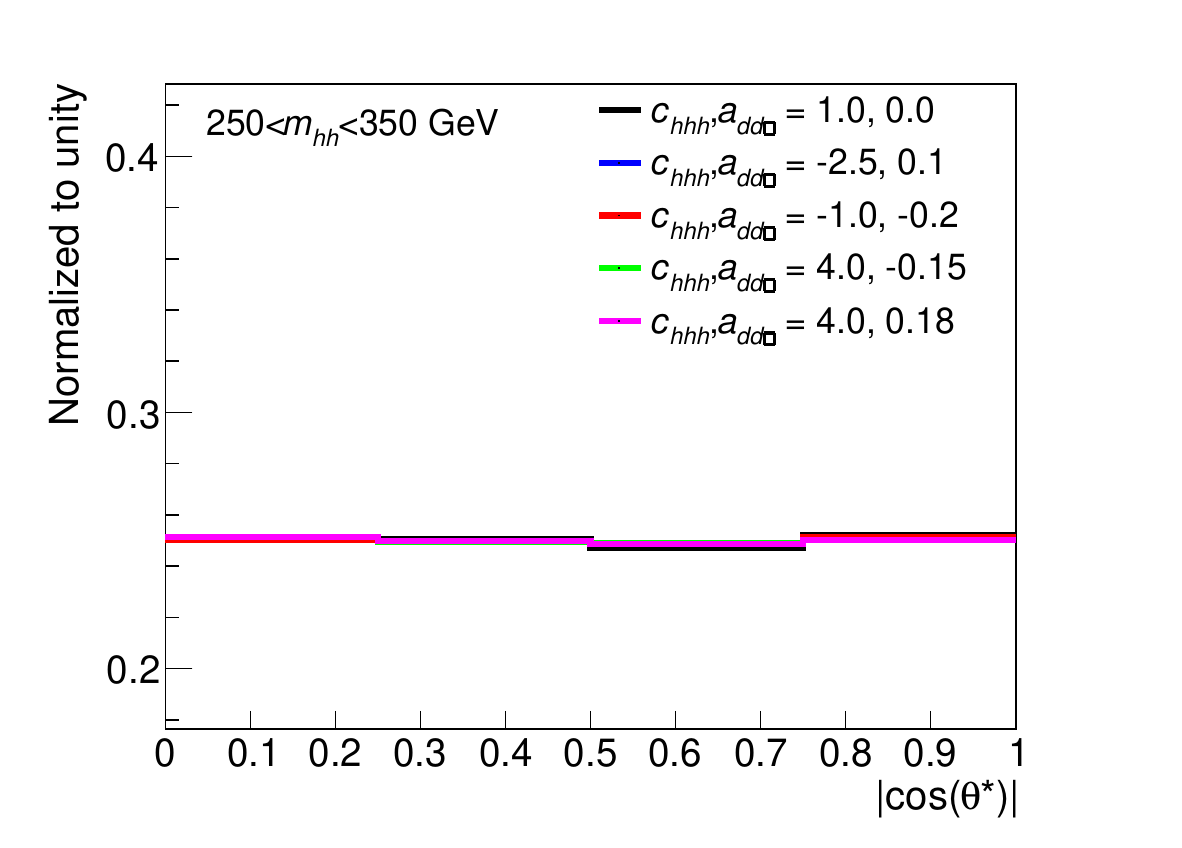} }
\hfill
\parbox{0.488\textwidth}{
\vspace{-3cm}
\caption{Representative distributions of the $h$ scattering angle for different values of the (phenomenology-dominating) HEFT coefficients of Eq.~\eqref{eq:rescale}, similar to Fig.~\ref{fig:mhhdists}. Here we focus on the threshold region of the $hh$ system that drives inclusive sensitivity. No discernible variation is visible as a function of the HEFT coefficients considered in this work.\label{fig:cossdistsLM}}}
\end{center}
\end{figure}

\begin{figure}[!t]
\begin{center}
\subfigure[]{\includegraphics[width=0.496\textwidth]{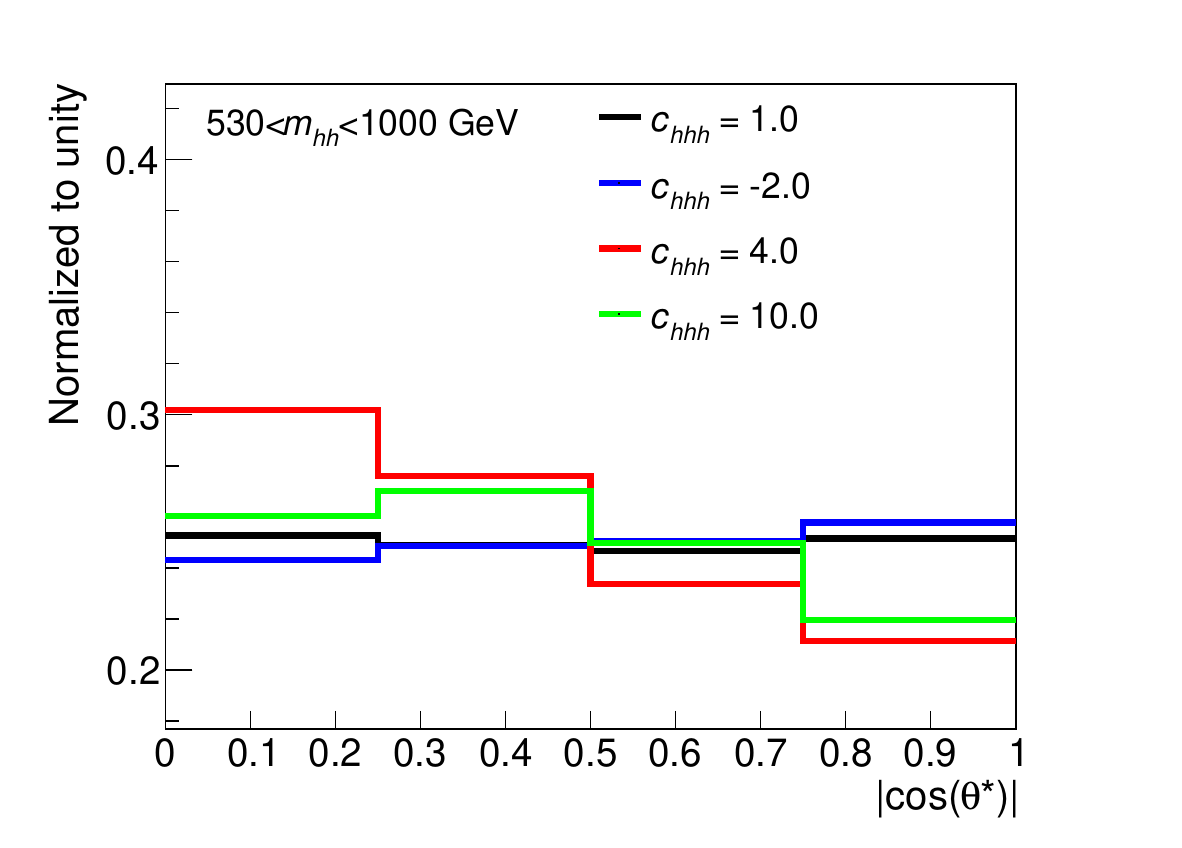}}\hfill
\subfigure[]{\includegraphics[width=0.496\textwidth]{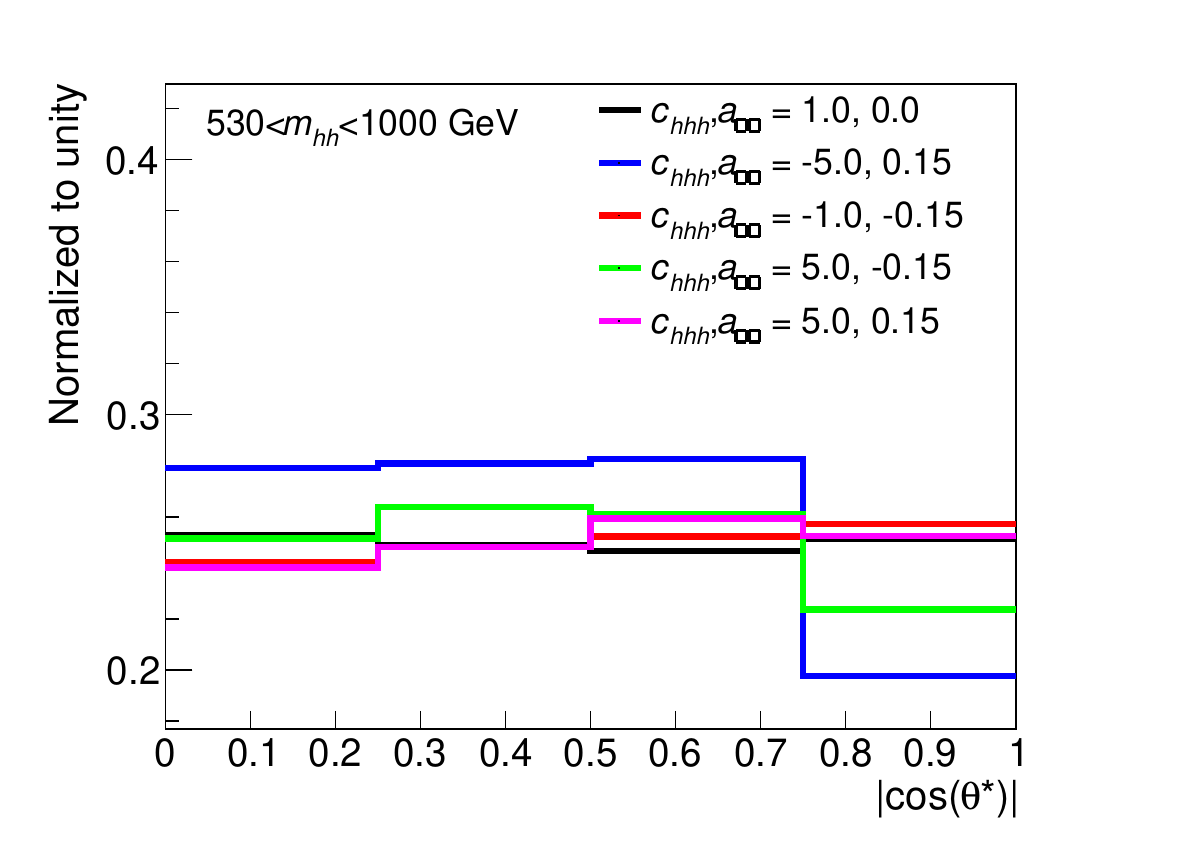}}
\subfigure[]{\includegraphics[width=0.496\textwidth]{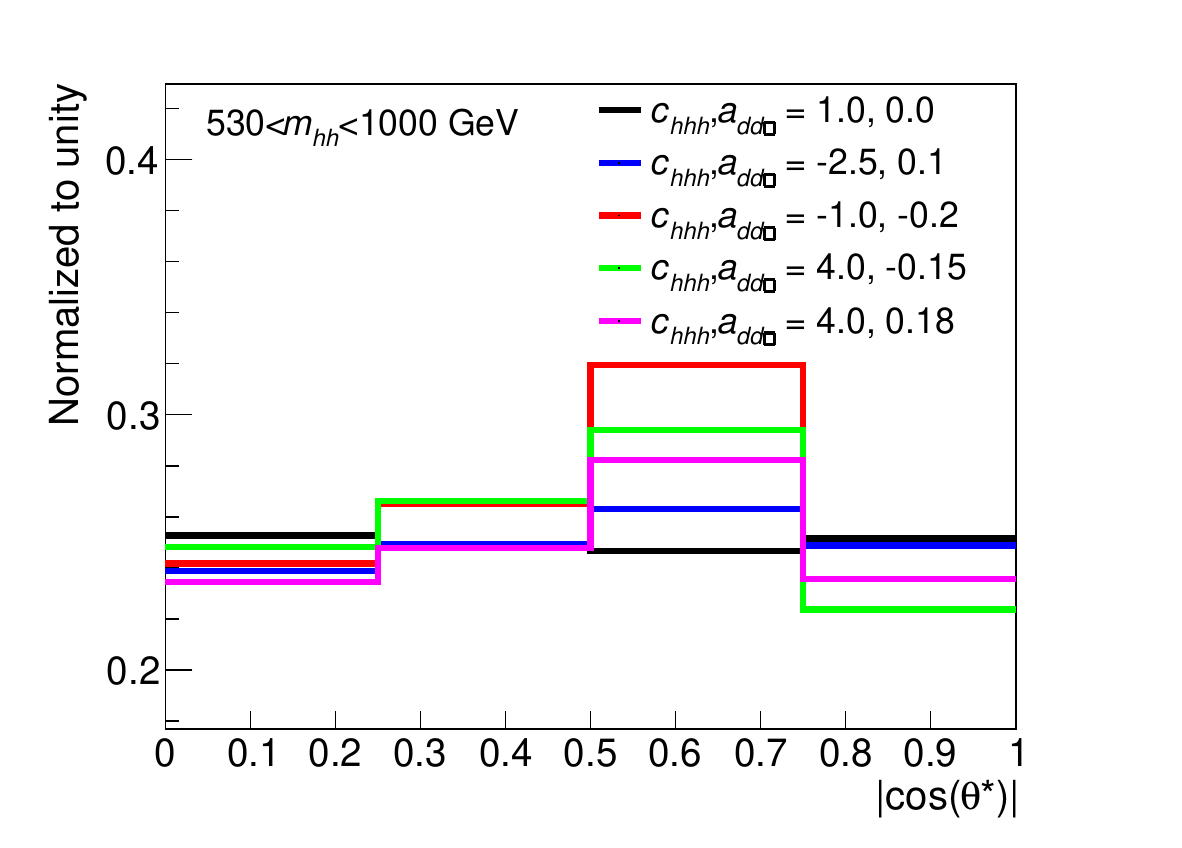} }
\hfill
\parbox{0.488\textwidth}{\vspace{-3cm}
\caption{Identical to Fig.~\ref{fig:cossdistsLM}, but focusing on the boosted $h$ region at large invariant di-Higgs masses. Correlations probe the interplay of the triangle diagrams with the box topologies for $gg\to hh$ production at one-loop, thus providing additional sensitivity that can be exploited in a multi-dimensional likelihood approach.\label{fig:cosdistsHM}}}
\end{center}
\end{figure}

Focussing on $b\bar b \gamma \gamma$ first, we base our analysis on the results of Ref.~\cite{ATL-PHYS-PUB-2025-001}. We have validated\footnote{As part of this validation, we have regenerated the dominant $b\bar{b} \gamma \gamma$ continuum background with \textsc{MadGraph5\_aMC@NLO}~\cite{Alwall:2014hca}, rescaled to have the same signal yield as the ATLAS result a function of $m_{hh}$ as given in the paper after all selections. This enables us to go beyond $m_{hh}$ as a single discriminant, see below.} our results against the reported Run 2 results, on which~\cite{ATL-PHYS-PUB-2025-001} itself builds as detailed in Ref.~\cite{aad2024studies}. A flat signal efficiency, computed as the average of the signal efficiency for SM and $\kappa_\lambda = 10$, is applied. These points for evaluating the signal efficiency are chosen because they are provided by ATLAS in Refs.~\cite{ATL-PHYS-PUB-2022-001,aad2024studies}, and their adequacy is evaluated by reproducing the reported limits. Further, we consider systematic uncertainties of 7\% and 10\% related to the background. These values are chosen as they reproduce the ATLAS Run 2 and HL projections and are consistent with the uncertainties quoted in the corresponding data tables of~\cite{ATL-PHYS-PUB-2022-001,PhysRevD.106.052001}.

Our analysis framework  is binned with two bins as $m_{hh} \in [250,350,1400]$~GeV and two bins in $|\cos{\theta^*}|$ with edges [0, 0.5, 1.0]. The bins in $|\cos{\theta^*}|$ are considered to enable better compatibility with the limits from ATLAS when performing the validations; however, they do not have a big impact beyond achieving a better validation of the inclusive selection. For the HL $b\bar b \gamma \gamma$ projections, the cross-section of the background and the signal is scaled by a factor 1.18 according to Ref.~\cite{ATL-PHYS-PUB-2022-001}, in accordance with LHCWG recommendations.

Turning to $b\bar b b\bar b$, we again validate our analysis against Ref.~\cite{PhysRevD.108.052003}, reproducing the relevant Run 2 results. We also performed cross-checks against Ref.~\cite{ATL-PHYS-PUB-2022-053}. Similar to $b\bar b\gamma \gamma$, we consider a two-dimensional likelihood build from binned $m_{hh}$ and $|\Delta\eta_{hh}|$ distributions.\footnote{The variable $X_{hh}$ used in \cite{PhysRevD.108.052003} is omitted because it is a reconstruction-level quantity not suitable for the analysis presented here.} A flat signal efficiency, calculated based on the average SM and for $\kappa_\lambda = 6$ signal efficiency, is applied, similar to $b\bar b \gamma \gamma$. We apply a flat signal efficiency, averaging the SM and $\kappa_\lambda = 6$ efficiencies as detailed above. Systematic uncertainties of 1\% or 2.5\% are assigned to the background. As for our choices in the $b\bar b \gamma \gamma$ channel, these values reproduce the ATLAS Run 2 and HL projections. For the HL-LHC signal and background are rescaled by 1.18~\cite{ATL-PHYS-PUB-2022-053}.

Throughout this work, we obtain limits from a likelihood-ratio test based on Wilks’ theorem~\cite{Wilks:1938dza}. The likelihoods are assumed to be Gaussian, and the histogram bins are treated as uncorrelated. This approach largely emulates procedures on the experimental side (see, e.g.,~\cite{aad2024studies,PhysRevD.108.052003} for additional information and discussion). For this work, we will limit ourselves to reporting the results of the statistical analysis.

\begin{figure}[!t]
\begin{center}
\subfigure[]{\includegraphics[width=0.496\textwidth]{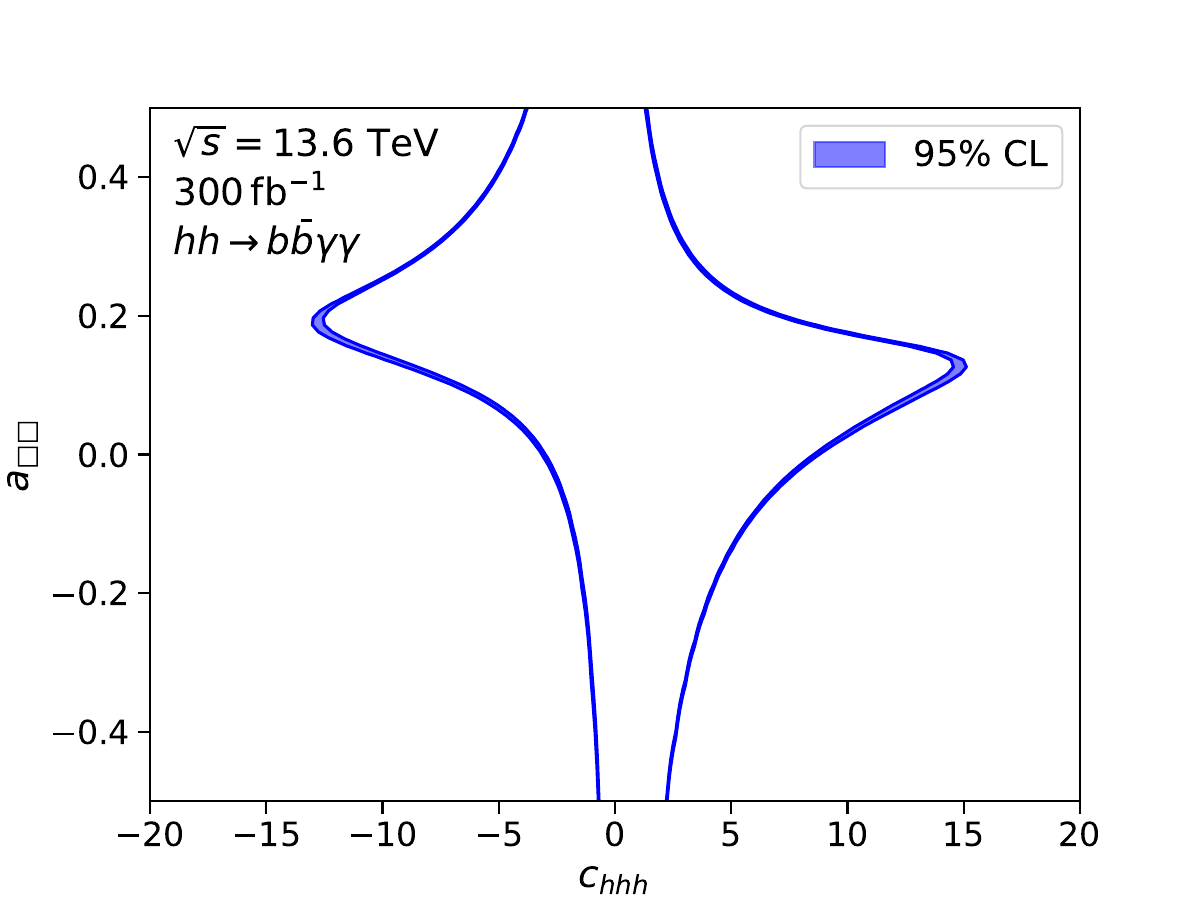}} \hfill 
\subfigure[]{\includegraphics[width=0.496\textwidth]{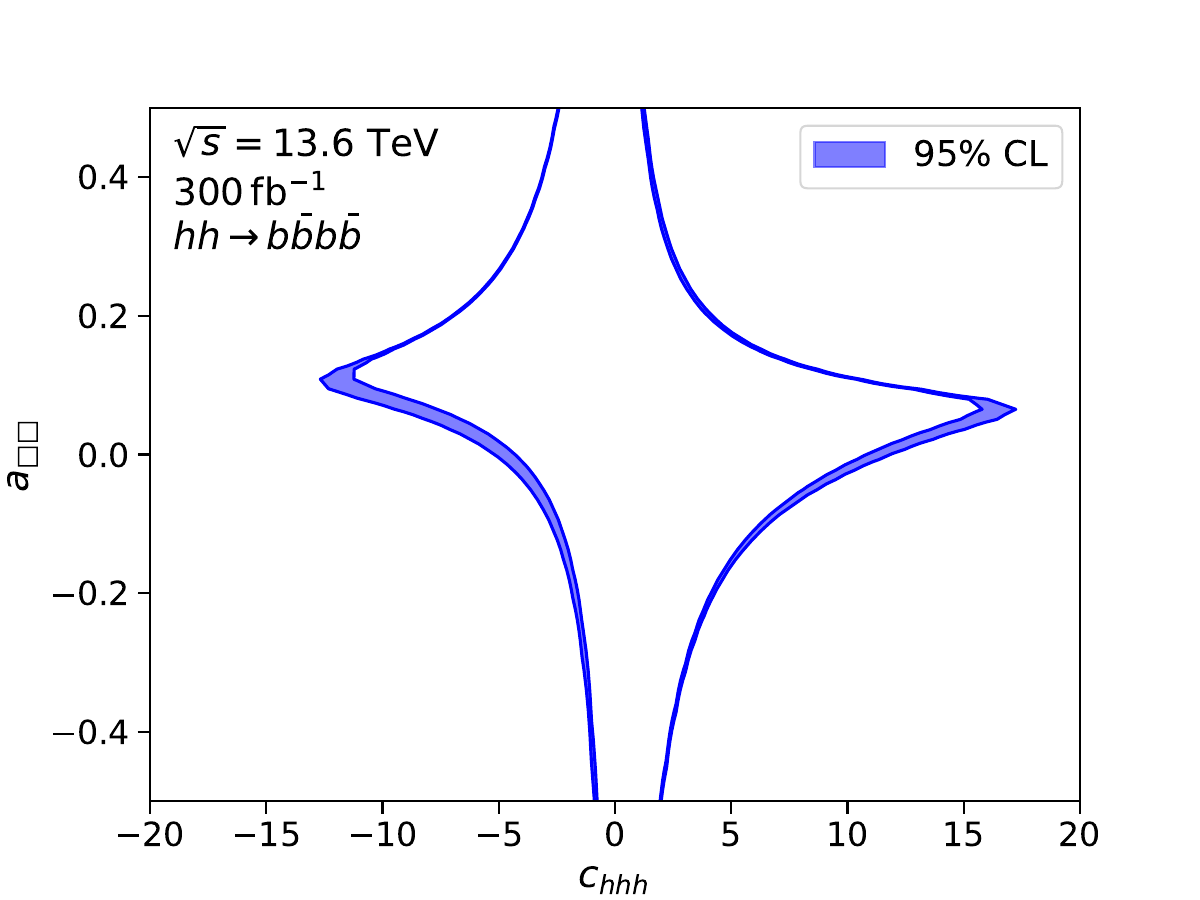}}
\subfigure[]{\includegraphics[width=0.496\textwidth]{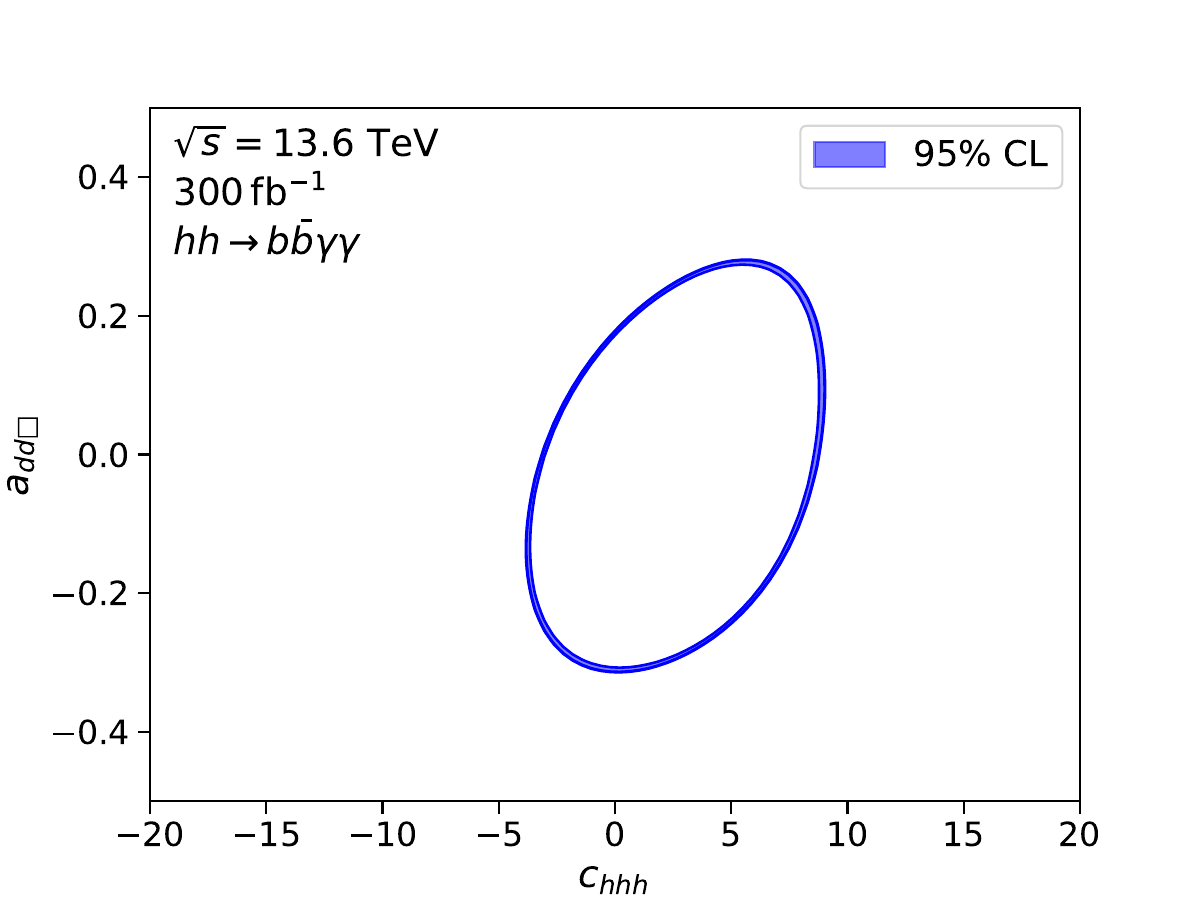}}\hfill
\subfigure[]{\includegraphics[width=0.496\textwidth]{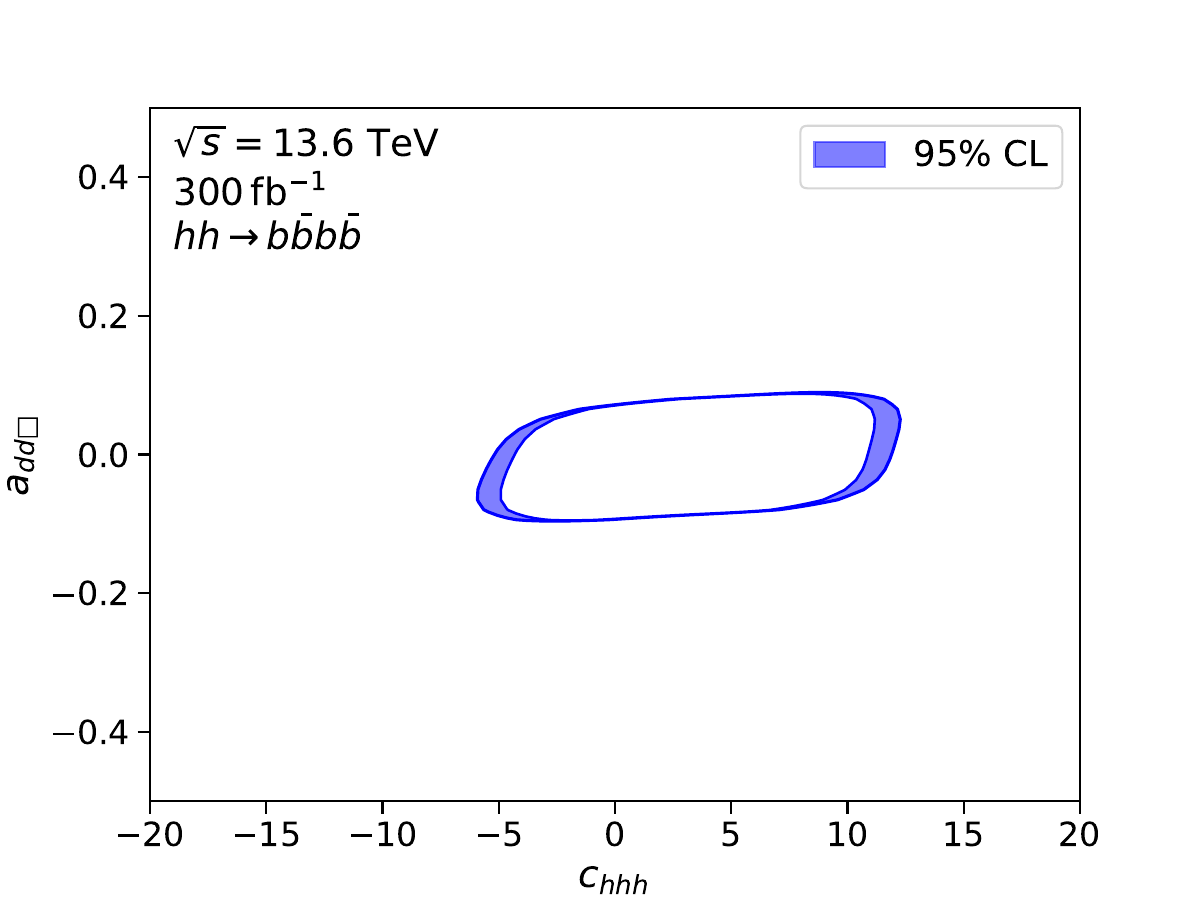}}
\caption{Expected exclusion contours resulting from our analysis for the LHC Run 3 in the $c_{hhh}$ and $a_{\Box\Box},a_{dd\Box}$ planes, respectively, for inclusive $b\bar b \gamma\gamma$ (including systematics ranging from 7\% to 10\%) (a,c) and $b \bar b b \bar b$ (systematics varying from 1\% for 2.5\% (b,d) (for details see the main text).\label{fig:run3}}
\end{center}
\end{figure}

As the bosonic HEFT can be implemented technically via a replacement of $c_{hhh}$, we can expect approximate cancellations when we compare momentum-dependent modifications with momentum-independent ones. Depending on the search channel, the event count is characterised by an average invariant di-Higgs mass, which can then lead to cancellations in the limit setting. We demonstrate this in Fig.~\ref{fig:run3}. In particular, $a_{\Box\Box}$ can be cancelled by an appropriate choice of $c_{hhh}$, leaving a wide area unconstrained when $hh$ is considered in isolation at Run 3. The relatively enhanced momentum dependence imparted by $a_{dd\Box}$ makes such cancellations less efficient. As expected, the harder $b\bar b b\bar b$ selection compared to the more inclusive $b\bar b \gamma \gamma$ makes it particularly effective in setting limits on momentum-enhanced contributions. This comes at the price of a slightly higher impact from systematics, which we represent by the width of the exclusion contour in Fig.~\ref{fig:run3}.

\begin{figure}[!t]
\begin{center}
\subfigure[]{\includegraphics[width=0.496\textwidth]{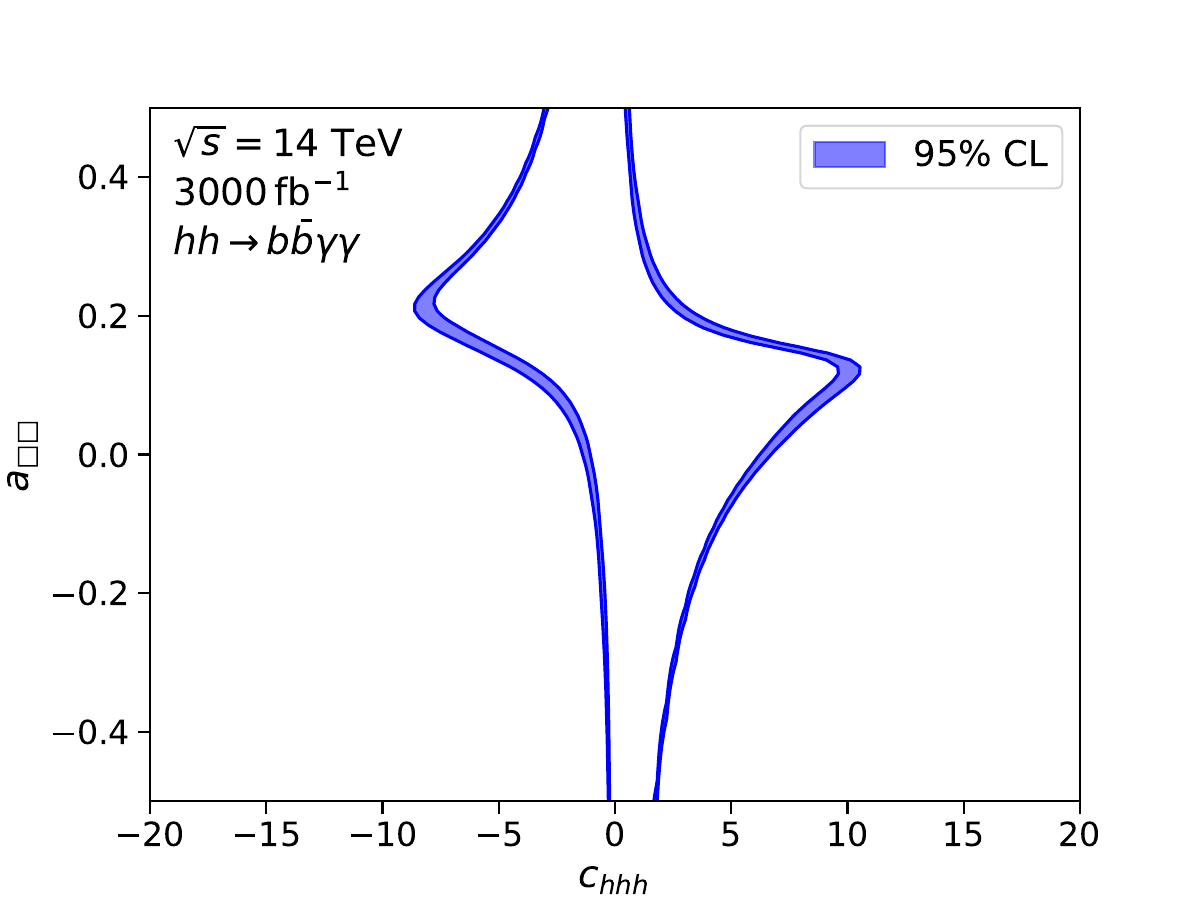}}\hfill
\subfigure[]{\includegraphics[width=0.496\textwidth]{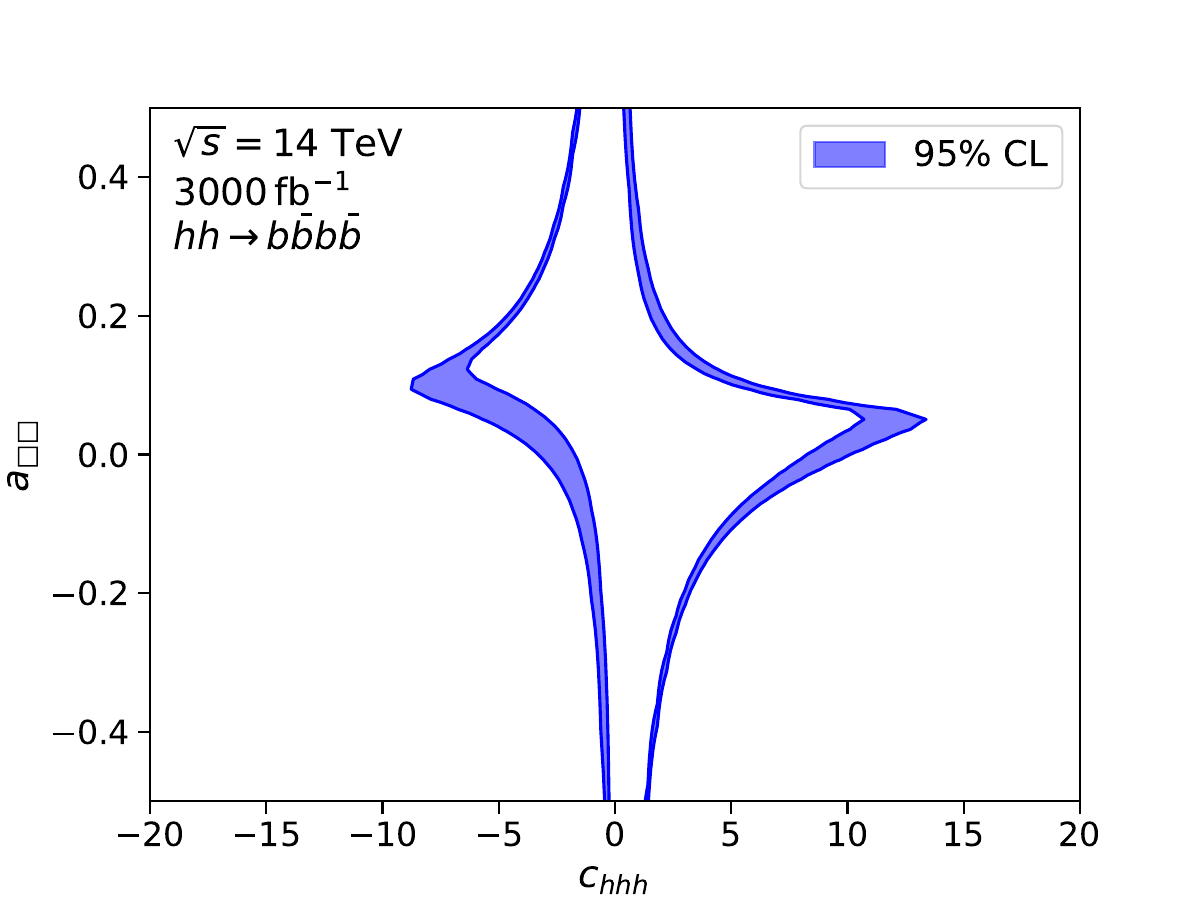}}
\subfigure[]{\includegraphics[width=0.496\textwidth]{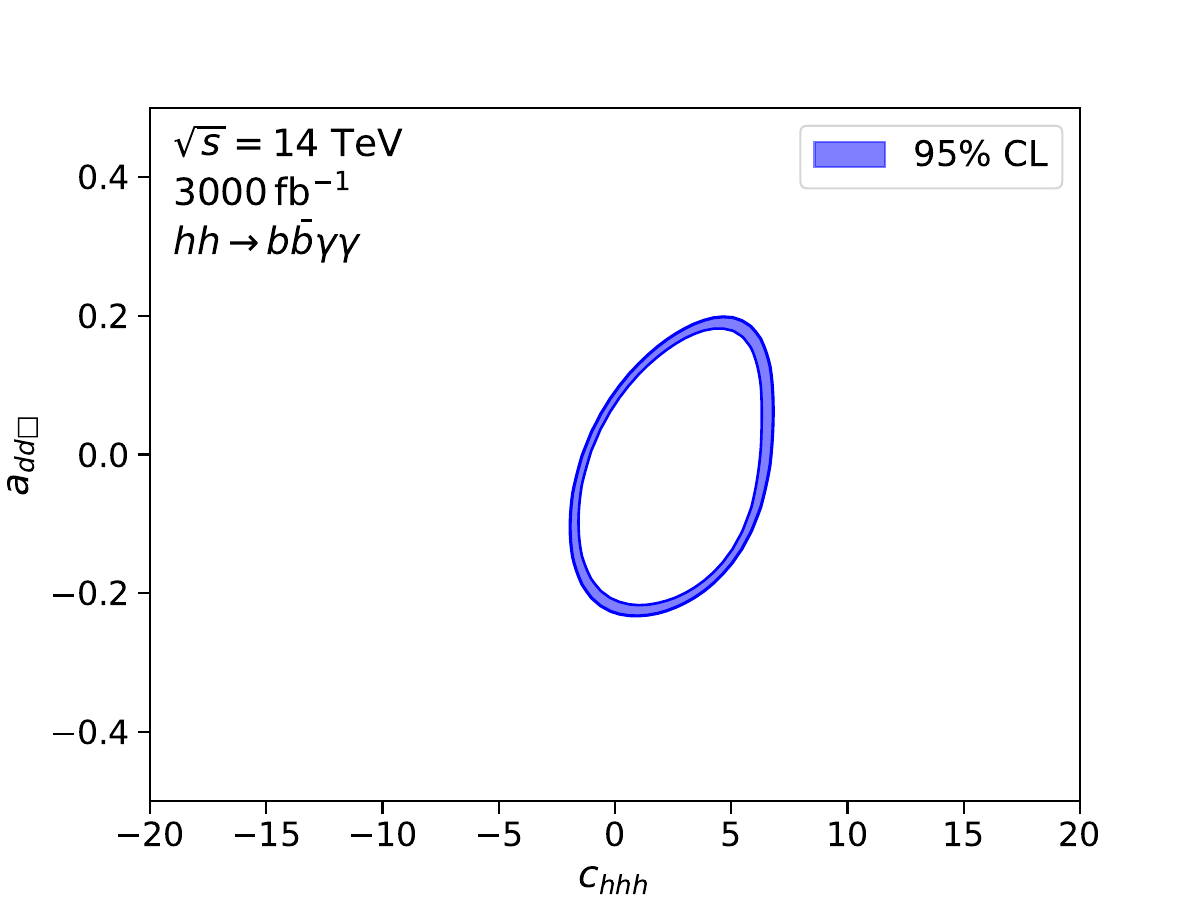}}\hfill
\subfigure[]{\includegraphics[width=0.496\textwidth]{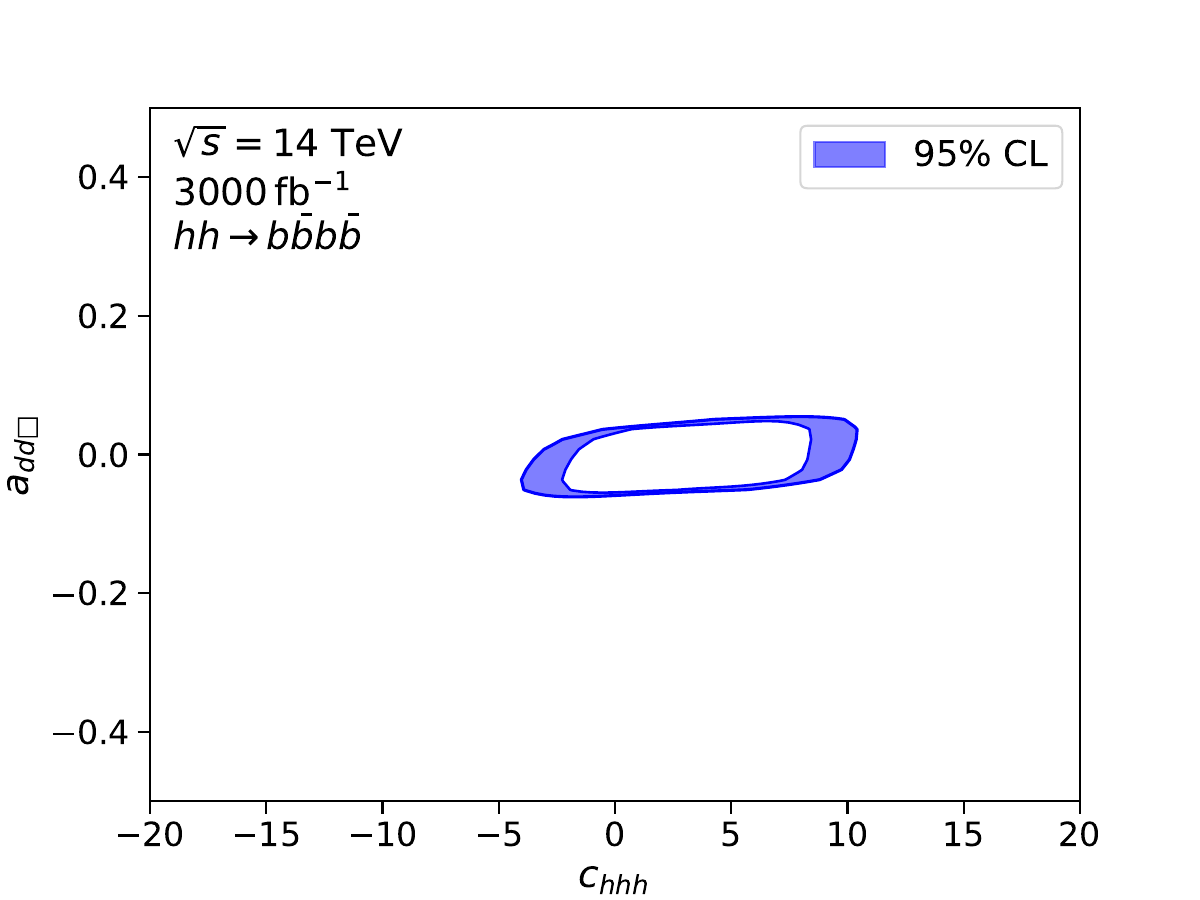}}
\caption{Expected exclusion contours resulting from our analysis for the HL-LHC in the $c_{hhh}$ and $a_{\Box\Box},a_{dd\Box}$ planes, respectively, for inclusive $b\bar b \gamma\gamma$ (including systematics ranging from 7\% to 10\%) (a,c) and $b \bar b b \bar b$ (systematics varying from 1\% for 2.5\% (b,d) (for details see the main text).\label{fig:hllhc}}
\end{center}
\end{figure}

The systematics do limit the sensitivity somewhat, but statistical uncertainties play a more dominant role for Run 3. This also becomes apparent from Tabs.~\ref{tab:Run34b1D_limits} and \ref{tab:1dbbyy_Run3_limits}, which provide individual limits for comparison. The latter improves significantly when we turn to the HL-LHC phase, where we see a broad improvement in sensitivity Fig.~\ref{fig:hllhc}. Constraints will improve by factors of up to two, particularly highlighting the boosted regime accessed in $b\bar b b\bar b$ that becomes under better statistical control for the $3/\text{ab}$ data set.

\begin{table}[!t]
  \centering
   \parbox{0.465\textwidth}{
  \centering
  \begin{tabular}{|l|c|c|}
    \hline
    \textbf{Coefficient} &  \textbf{95\% CL Limit}       & $\boldsymbol{\sigma_{\rm syst}}$ \\ 
    \hline
    $a_{\Box\Box}$              & [$-1.17,0.50$]        & 7\%   \\ 
    $a_{\Box\Box}$              & [$-1.19,0.52$]          & 10\%  \\ 
    \hline
    $a_{dd\Box}$               & [$-0.27,0.17$]         & 7\%   \\ 
    $a_{dd\Box}$               & [$-0.28,0.17$]         & 10\%  \\ 
    \hline
  \end{tabular}
  \caption{Run 3 one‐dimensional 95\% CL limits and systematic uncertainties for $b\Bar{b}\gamma\gamma$.}
  \label{tab:1dbbyy_Run3_limits}
   }\hfill
     \parbox{0.465\textwidth}{
  \centering
  \begin{tabular}{|l|c|c|}
    \hline
    \textbf{Coefficient} & \textbf{95\% CL Limit} & $\boldsymbol{\sigma_{\rm syst}}$ \\ 
    \hline
    $a_{\Box\Box}$ & [$-0.97,0.50$] & 1\%  \\ 
    $a_{\Box\Box}$ & [$-0.99,0.54$] & 2.5\%\\ 
    \hline
    $a_{dd\Box}$  & [$-0.081,0.063$] & 1\%   \\ 
    $a_{dd\Box}$  & [$-0.081,0.063$] & 2.5\% \\ 
    \hline
  \end{tabular}
  \caption{Run 3 one‐dimensional 95\% CL limits and systematic uncertainties for $b\Bar{b} b\Bar{b}$.}
  \label{tab:Run34b1D_limits}}
\end{table}

\begin{table}[!t]
  \centering
    \parbox{0.465\textwidth}{
  \centering
  \begin{tabular}{|l|c|c|}
    \hline
    \textbf{Coefficient} &  \textbf{95\% CL Limit}       & $\boldsymbol{\sigma_{\rm syst}}$ \\ 
    \hline
    $a_{\Box\Box}$              & [$-0.91,0.26$]         & 7\%   \\ 
    $a_{\Box\Box}$              & [$-0.96,0.31$]         & 10\%  \\ 
    \hline
    $a_{dd\Box}$               & [$-0.20,0.10$]         & 7\%   \\ 
    $a_{dd\Box}$               & [$-0.21,0.11$]        & 10\%  \\ 
    \hline
  \end{tabular}
  \caption{HL one‐dimensional 95\% CL limits and systematic uncertainties for $b\Bar{b}\gamma\gamma$.}
  \label{tab:1dbbyy_HL_limits}
}
\hfill
\parbox{0.465\textwidth}{
  \centering
  \begin{tabular}{|l|c|c|}
    \hline
    \textbf{Coefficient} & \textbf{95\% CL Limit} & $\boldsymbol{\sigma_{\rm syst}}$ \\ 
    \hline
    $a_{\Box\Box}$ & [$-0.63,0.23$] & 1\%  \\ 
    $a_{\Box\Box}$ & [$-0.67,0.32$] & 2.5\%\\ 
    \hline
    $a_{dd\Box}$  & [$-0.045,0.025$] & 1\%   \\ 
    $a_{dd\Box}$  & [$-0.045,0.025$] & 2.5\% \\ 
    \hline
  \end{tabular}
  \caption{HL one‐dimensional 95\% CL limits and systematic uncertainties for $b\Bar{b} b\Bar{b}$.}
  \label{tab:HL4b1D_limits}}
\end{table}

\subsection{Relevance of four top production}
\label{sec:4top}
As the bosonic HEFT introduces considerably more parameters to the $hh$ final states, additional complementary information needs to be included to tighten constraints. With $a_{\Box\Box}$ modifying Higgs propagation off-shell, two LHC processes are singled out, where such a behaviour is observable. Firstly, in $gg\to ZZ$, the Higgs contribution remains sizable even for large $ZZ$ invariant masses due to unitarity cancellations expected in the SM~\cite{Kauer:2012hd,Englert:2014aca}. This provides an avenue for setting constraints on $a_{\Box\Box}$ as discussed recently in~\cite{Anisha:2024xxc}. Notably, momentum enhancements of a similar type in SMEFT cancel to leading order in this final state~\cite{Englert:2019zmt}. The second process is four top production, which is particularly sensitive to electroweak contributions~\cite{Frederix:2017wme} and has been highlighted as a versatile tool for BSM discovery~\cite{Alvarez:2016nrz,Darme:2018dvz,Alvarez:2019uxp,Darme:2021gtt,Blekman:2022jag,Anisha:2023xmh}. We can expect four top production to reach good sensitivity towards the HL-LHC phase as it is currently systematics limited with experiments displaying impressive progress in enhancing their sensitivity beyond naive extrapolations (see also~\cite{Belvedere:2024wzg}). Hence, we expect to reach very good sensitivity to $a_{\Box\Box}$ when these processes are scrutinised from this angle, and we will consider their interplay with the $gg\to hh$ discussion above in more detail.

\begin{figure}[!t]
\begin{center}
\subfigure[]{\includegraphics[width=0.496\textwidth]{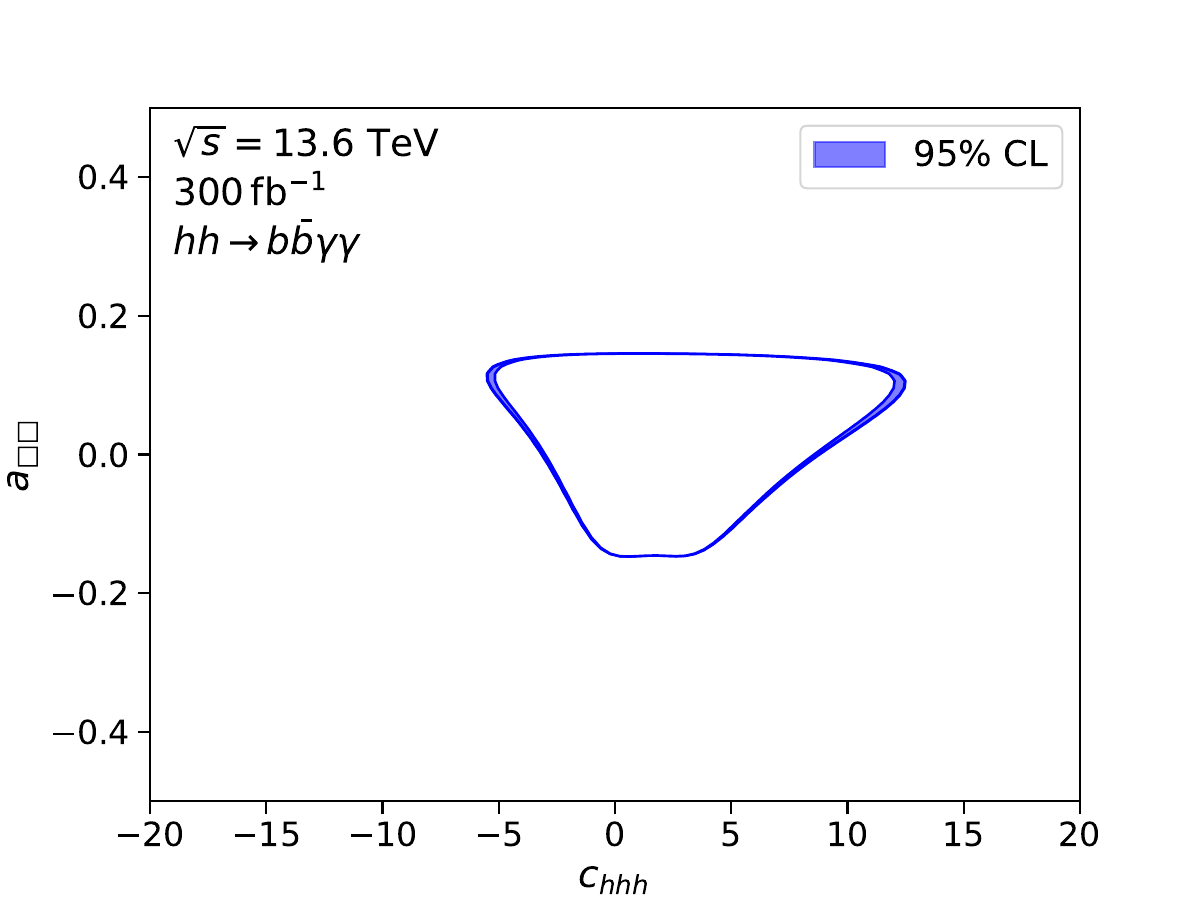}}\hfill
\subfigure[]{\includegraphics[width=0.496\textwidth]{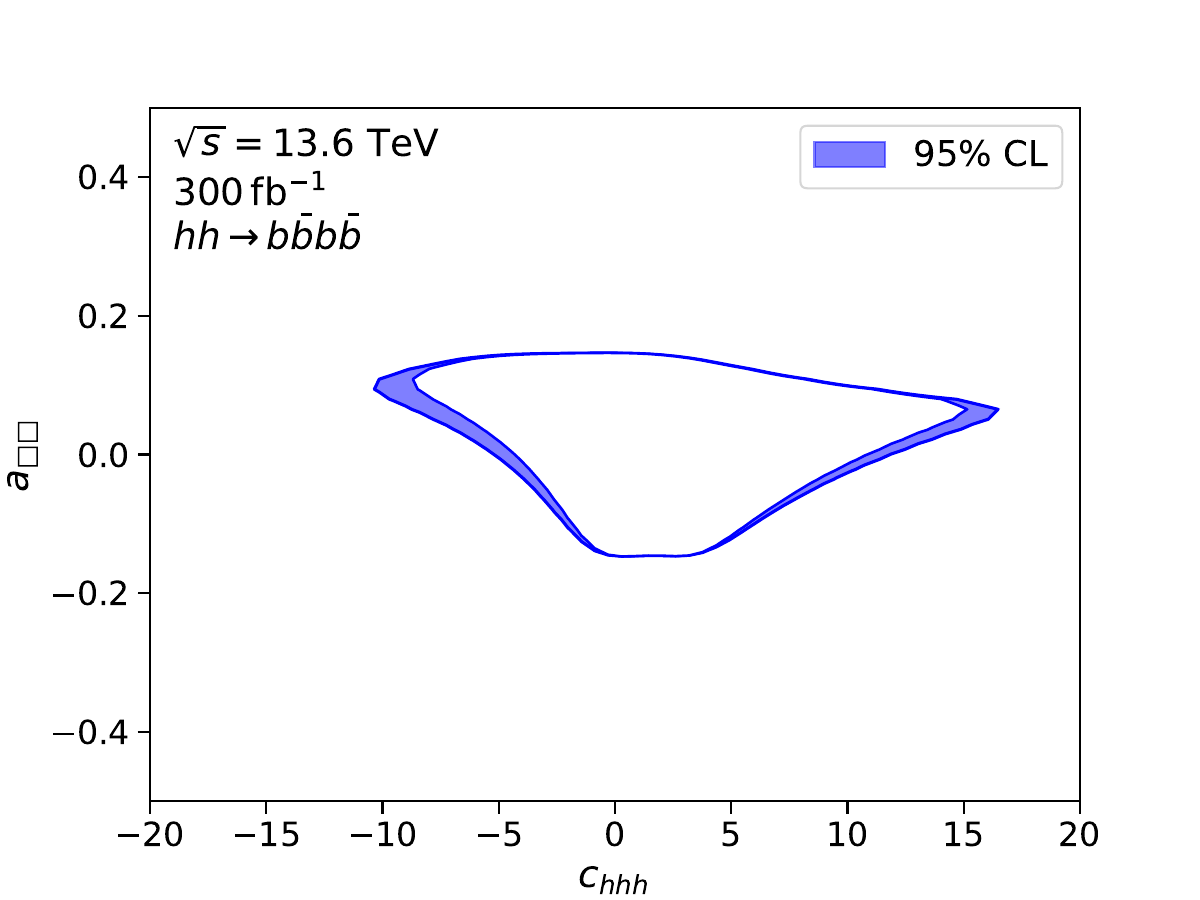}}
\caption{Similar to Fig.~\ref{fig:run3}, but including constraints from four top measurements at Run 3. The external constraint from the four tops on $a_{\Box\Box}$ is included as a Gaussian constraint with $\sigma = 0.06$ centred at zero.\label{fig:4tr3}}
\end{center}
\end{figure}

\begin{figure}[!t]
\begin{center}
\subfigure[]{\includegraphics[width=0.496\textwidth]{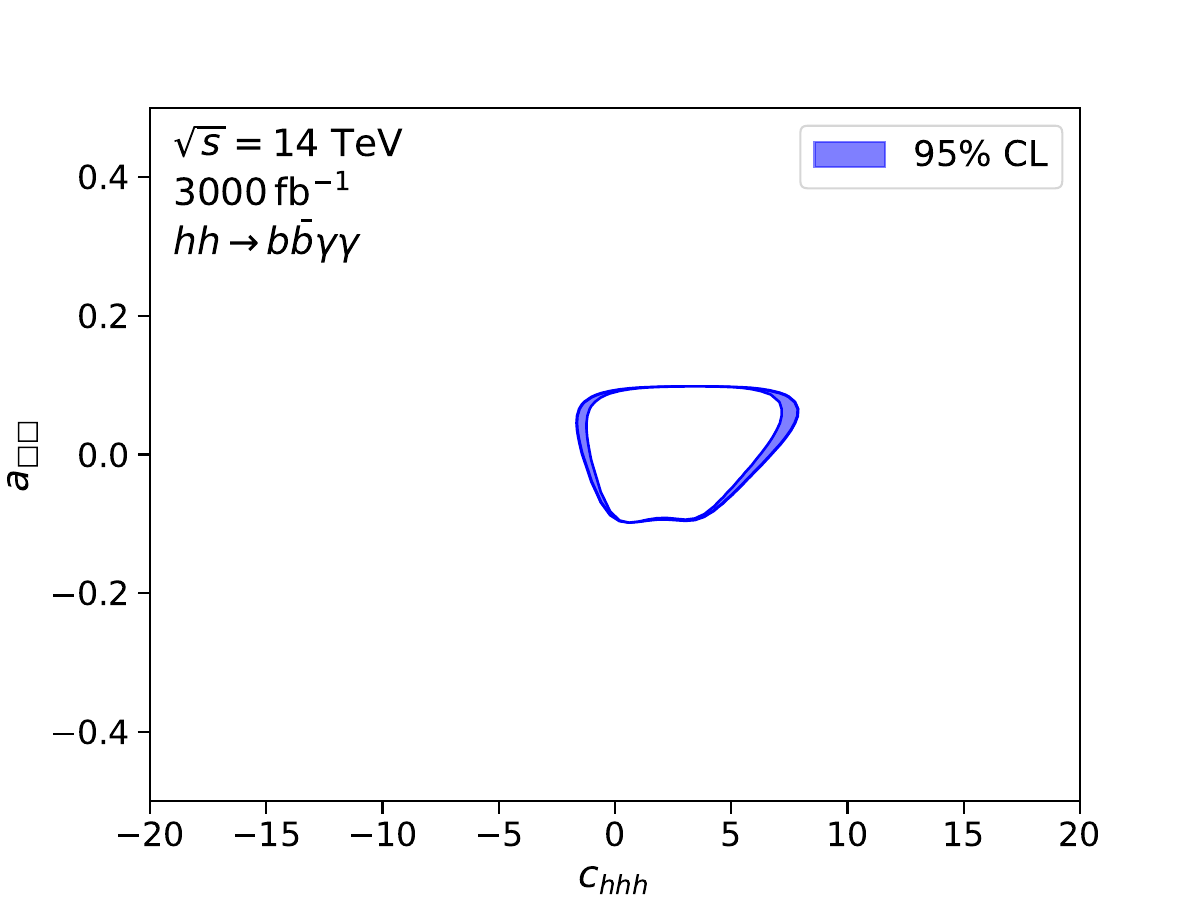}}\hfill
\subfigure[]{\includegraphics[width=0.496\textwidth]{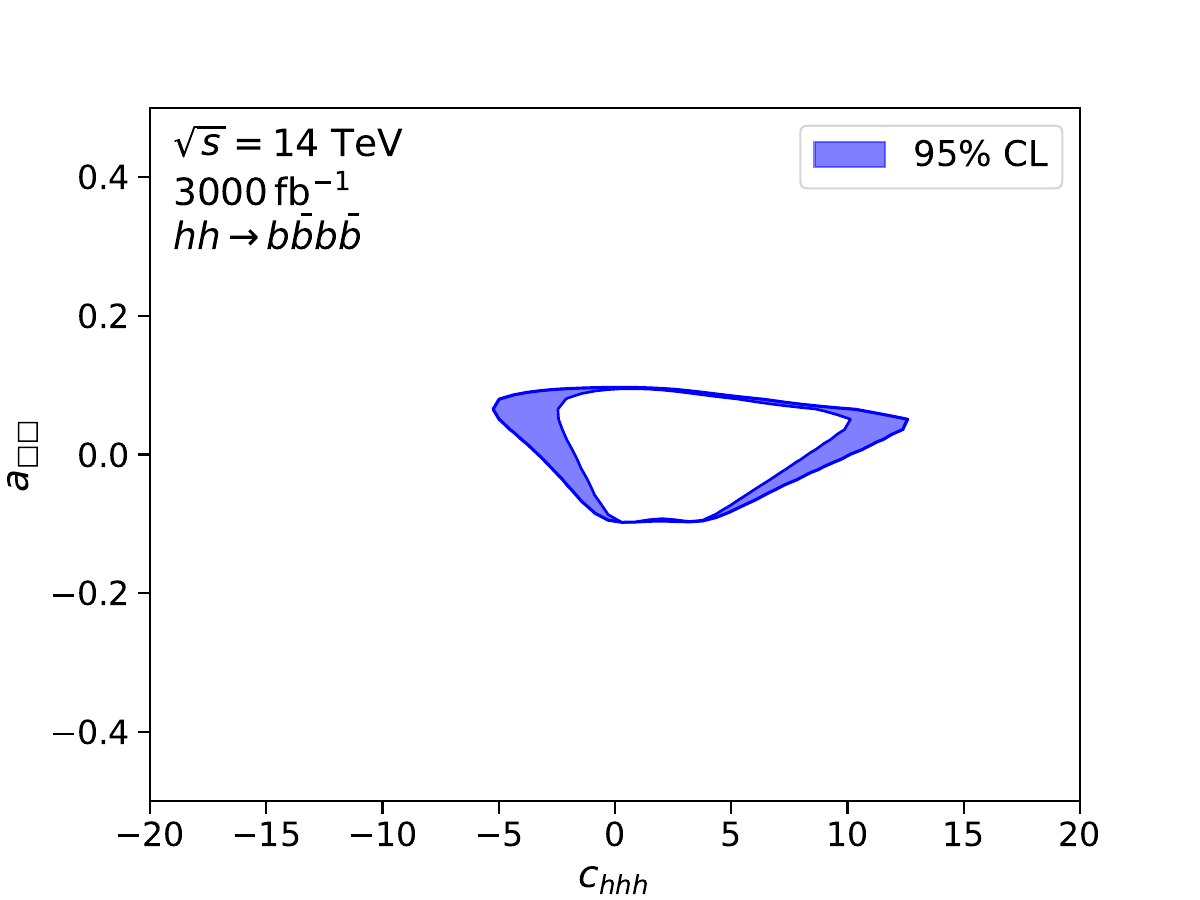}}
\caption{Similar as Fig.~\ref{fig:hllhc}, but including constraints from four top measurements at HL. The external constraint from four tops on $a_{\Box\Box}$ is included as a Gaussian constraint with $\sigma = 0.04$ centered at zero.\label{fig:4thllc}}
\end{center}
\end{figure}

Obtaining the four top estimate using \textsc{MadGraph5\_aMC@NLO}~\cite{Alwall:2014hca} through a modified {\sc{Helas}} implementation~\cite{Murayama:1992gi} that evaluates the Higgs two-point function in the on-shell scheme~\cite{Herrero:2021iqt,Anisha:2024ryj}. Our analysis of four top quark production is based on Ref.~\cite{CMS:2018nqq}, and we combine $hh$ and $t\bar t t\bar t$ final states to obtain constraints in the $c_{hhh},a_{\Box\Box}$ plane, Figs.~\ref{fig:4tr3} and~\ref{fig:4thllc}. Both at the LHC Run 3, as well as for the HL-LHC phase we see good sensitivity improvements compared to $hh$ in isolation: The vertical $a_{\Box\Box}$ direction gets tightly constrained even for the $b\bar b \gamma \gamma$ selection, where such constraints are not achievable in $hh$ alone at the HL-LHC.\footnote{It is worthwhile pointing out that the constraints obtained this way on $a_{\Box\Box}$ are tighter than the one obtained from $gg\to ZZ$~\cite{Anisha:2024xxc}.} This highlights four top production not only as a very relevant channel in determining more general deviations from the SM than modelled by SMEFT at dimension 6, but also its interplay in setting constraints on more general new physics models when considered in concert with other processes that will unfold their full statistical power the HL phase, such as $hh$ production.

\section{Summary and Conclusions}
\label{sec:conc}
Di-Higgs final states, particularly in the dominant gluon fusion production mode, remain a central target for probing the structure of the Higgs sector and searching for new physics beyond the Standard Model. While SMEFT and $\kappa$-framework approaches have guided most current experimental strategies, these frameworks impose rigid correlations between Higgs interactions across different multiplicities. In contrast, HEFT allows for more general, momentum-dependent deformations that can decouple di-Higgs observables from single-Higgs measurements and thereby offer a richer phenomenological landscape.

In this work, we proposed and implemented a momentum-dependent reweighting scheme that enables the inclusion of bosonic HEFT interactions up to chiral dimension four into existing Monte Carlo workflows. This strategy allows for the reinterpretation of current and future di-Higgs searches in a more general EFT framework, without requiring significant changes to experimental analyses. We showed that HEFT operators, particularly $a_{\Box\Box}$ and $a_{{dd}\Box}$, can introduce substantial shape modifications to the di-Higgs invariant mass distribution $m_{hh}$, especially in boosted regimes. These modifications can result in differential sensitivities across final states such as $b\bar{b}b\bar{b}$ and $b\bar{b}\gamma\gamma$, which we analysed using a multidimensional likelihood approach.

For LHC Run 3, we provided one-dimensional 95\% confidence level bounds on key HEFT coefficients, demonstrating that momentum-enhanced operators like $a_{dd\Box}$ can be effectively constrained in channels with harder event selections. In contrast, $a_{\Box\Box}$, which can partially mimic standard $\kappa$-type modifications, remains more challenging to isolate in inclusive channels unless accompanied by shape-sensitive observables. These challenges are alleviated at the HL-LHC, where improved statistical precision and access to boosted kinematic regimes—particularly in the $b\bar{b}b\bar{b}$ final state—lead to significantly tighter bounds. For example, the HL-LHC can constrain $a_{dd\Box}$ at the per cent level, highlighting the potential to resolve non-linear dynamics in the Higgs sector.

Moreover, we emphasised the value of combining di-Higgs measurements with other rare LHC final states. In particular, four-top production $pp \rightarrow t\bar{t}t\bar{t}$ provides sensitivity to HEFT coefficients such as $a_{\Box\Box}$, which enter via modified Higgs propagators. We showed that including current and projected four-top constraints significantly improves exclusion contours in the ($c_{hhh}$, $a_{\Box\Box}$) plane, even in channels where di-Higgs sensitivity is limited. This underscores the complementarity of rare processes in building a comprehensive picture of BSM Higgs interactions.

Thus, the study shows that general momentum-dependent effects encoded in HEFT can leave distinctive imprints in di-Higgs kinematics accessible with current experimental tools. Furthermore, we outline a strategy to explore non-linear Higgs dynamics at the LHC and beyond by utilising simple reweighting techniques and combining information across multiple final states and rare processes. Applying this strategy to a high-luminosity extrapolation, the LHC will provide a relatively detailed picture of non-linear effects in the Higgs sector, in particular when viewed in concert with other rare final states such as four top production that highlight off-shell Higgs properties.

\subsection*{Acknowledgements}
C.E. is supported by the STFC under grant ST/X000605/1, and by the Leverhulme Trust with a Research Fellowship RF-2024-300$\backslash$9. C.E. is further supported by the Institute for Particle Physics Phenomenology Associateship Scheme. J.S. and T.I.C. are supported by the Swedish Research Council grant no. 2023-04654. M.S.~is supported by the STFC under grant ST/P001246/1. J.S. and T.I.C. acknowledge the use of the Fysikum HPC Cluster at Stockholm University.

\bibliography{paper.bbl}
\end{document}